\documentclass [10pt]{iopart}
\usepackage{epsf}
\usepackage{BoxedEPS}
\SetepsfEPSFSpecial 
\HideDisplacementBoxes

\begin{document}

\title{Charged-Current Disappearance Measurements in the NuMI Off-Axis Beam}

\author {R H Bernstein, Fermi National Accelerator Laboratory\footnote[1]{Fermi National Accelerator Laboratory, Batavia IL 60510 USA, rhbob@fnal.gov}}

\begin{abstract}{This article studies the potential of combining charged-current disappearance measurements of
$\nu_{\mu} \rightarrow
\nu_{\tau}$ from MINOS and an off-axis beam. I find that the error on $\Delta m^2$ from a 100 kt-yr off-axis measurement is a few percent of itself. Further,
I find little improvement to an off-axis measurement by combining it with MINOS.}\end{abstract}

 Several authors have suggested that an off-axis beam can be used in combination with the MINOS measurement to measure
$\Delta m^2$ in a NuMI off-axis beam to roughly 1\% with a 100 kt-yr exposure, {\it e.g.} \ $\Delta m^2 = .0030  \pm .00030$.\cite{velasco,para}  I
have examined this claim with a simulation using a standard off-axis configuration, approximate detector resolutions, and the Feldman-Cousins
prescription for the construction of $\Delta
\chi^2$. I find this claim unwarranted unless we assume $\sin^2 2  \theta =1$ and make optimistic assumptions concerning the errors.

\section {Physics and Detector Assumptions}

\begin{figure}[h]
\hglue 0.25in\BoxedEPSF{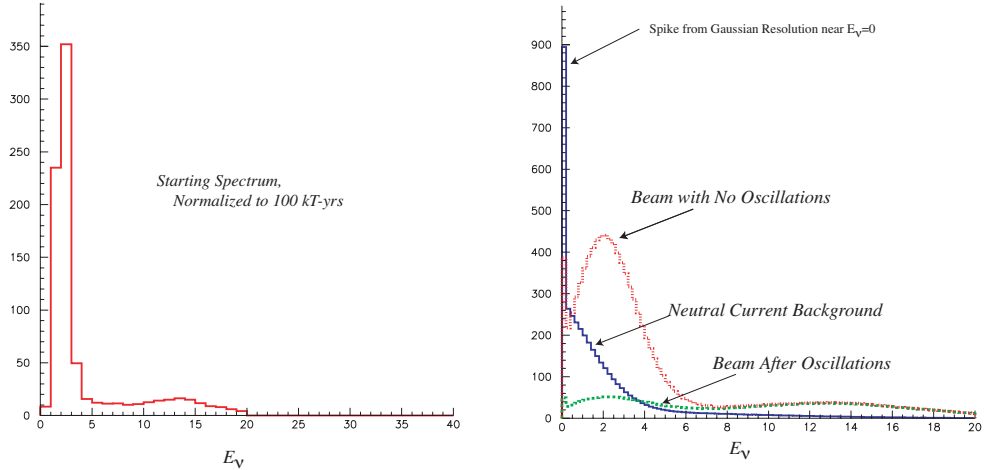 scaled 350}
\caption{ On the left is the spectrum for a 10 km off-axis detector before oscillations. On the right is the spectrum for signal and backgrounds. The spike at
the origin is from neutrinos smeared to ``negative" neutrino energy that I reconstruct in the zero energy bin.\label{fig:spectrum}}
\end{figure}

I assume an off-axis detector with $r=10$ km off the beam axis at the FNAL/Soudan distance of 732 km.  The spectrum before is applied is shown on the
left-hand side of Fig.~\ref{fig:spectrum} assuming {\it no} oscillations.  

The spectrum is sufficiently narrow that it is instructive to consider it to be a $\delta$-function.  Then  there is no spectral information since all
neutrinos are at the same energy, and only a total rate test can be performed.   In this case neutral current contamination with low missing neutrino energy
is the dominant source of background.  This analysis is based on the spectral test but this argument illustrates why the background is the dominant source of
error.

  Based on  the typical current detector designs I posit a non-magnetized detector.  There is little advantage to a mangetic field because
there is simply not enough lever arm to determine the muon momentum for curvature.  Calorimetry is performed by hit-counting, so that the total number of hits
is roughly proportional to neutrino energy.  No muon tracking is attempted.  Normally one would look for long tracks protruding past the end of the hadronic
shower to signal the presence of a muon, but at these energies all outgoing tracks are very close in length.   The error on momentum from  length
determination goes as a (constant term dependent upon straggling)/track length, and with the short track lengths at a GeV or less this is not sufficiently
precise. Hence the best way to determine momentum is from counting hits as a measurement of $dE/dx$ energy deposit.   Discussions with FMMF collaborators who
used hit counting give resolutions typically of order $1.0/\sqrt{E}$.\cite{hatcher}    In what follows I will assume 55\%/$\sqrt{E}$ for the shower as a
``best-possible" case.  

\section{Uncertainties}

The sources of uncertainty used in this study  are given in Table~\ref{tab:errtab}. I assume a 100 kTon$\cdot$yr exposure, or 10 years of calendar time for a
10 kTon detector. The error on the correlated flux could come from normalization of fiducial volume and extrapolation to the off-axis detector.  Random flux
errors are identical to those assumed by MINOS.  Errors on the shape of the extrapolated spectrum come from uncertainties in FLUKA or GEANT and the precise
location of beam elements.  The values for the shape uncertainty  are based on discussions with and work by Para and Szleper.\cite{szleper}

\vspace*{0.125in}
\begin{table}[h]
\begin{tabular}{|l|c|c|}\hline Statistical& 100 kton$\cdot$ years & \\\hline
\multicolumn{3}{c}{~~}\\\hline Beam& & \\\hline Correlated Flux& 3\% & \\ Random Flux& 2\% in any 1 GeV bin& \\ Shape& $A \sin (\lambda E_{\nu}/5. + \phi)$ & 
~~ \\
 & $ -.10 < A < .10$ flat&   See \\
 & $ 0< \lambda < 2\pi\times 5$ flat&    hep-ex/0110001, \\
  & $ 0 < \phi < 2 \pi$ flat&  0110032 \\\hline
\multicolumn{3}{c}{~~}\\\hline Detector& & \\\hline Hadronic Energy & 0.55/$\sqrt{E}$& \\ Muon Momentum& not separately seen& \\ & include with hadron shower
energy&\\\hline
\end{tabular} 
\caption{Sources of Uncertainty Assumed in this Analysis.\label{tab:errtab}}
\end{table}

\section{Results}
  The construction of the 90\% CL levels are made using the Neyman-Pearson construction, as re-invented by Feldman and Cousins.\cite{bob_and_gary}; this 
method correctly handles the $\sin^2 2\theta = 1$ physical boundary.

\begin{figure}
\begin{center}
\BoxedEPSF{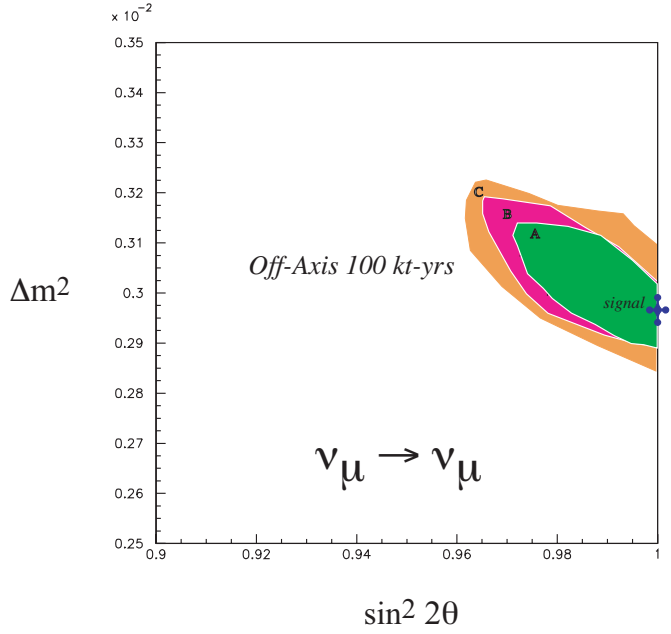 scaled 425}
\end{center}
\caption{Full set of analyzed errors: (A) Statistical and Detector Resolution, (B) Beam Rate and Spectrum Errors, and (C) Effect of Neutral Current
Background. The inner region is Set A, middle region is Sets A and B, outer region is Sets A, B, and C.  \label{fig:allerrors}}
\end{figure}

\begin{figure}
\begin{center}
\BoxedEPSF{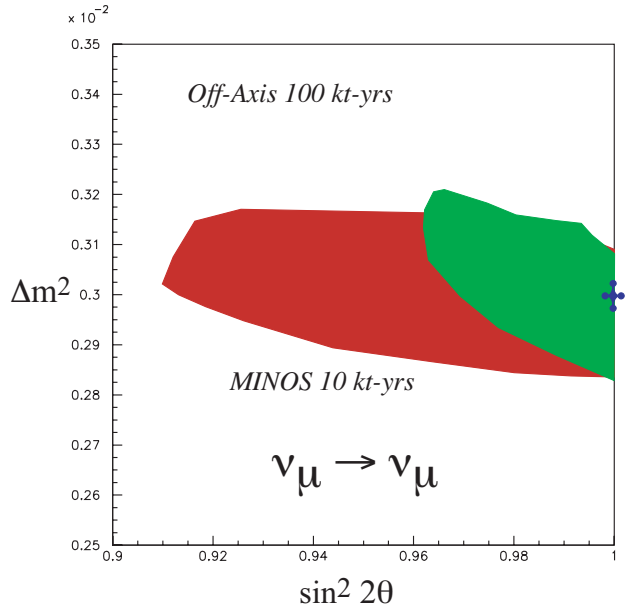 scaled 425 }
\end{center}
\caption{Comparison to MINOS capabilties, all at 90\% CL, for the full set of analyzed errors: (A) Statistical and Detector Resolution, (B) Beam Rate and
Spectrum Errors, and (C) Effect of Neutral Current Background. \label{fig:answer}}
\end{figure}

The results were extracted in three stages.  The first used statistical and resolution errors but assumed perfect beam knowledge.  The second added the
effects of the assumed beam uncertainties.  Finally,  Fig.~\ref{fig:allerrors} includes these errors and the result of the statistical fluctuations of the
neutral current background. I compare to the result for MINOS in Fig.~\ref{fig:answer}.  The errors are the standard MINOS errors and can be found in
Ref.\cite{mynuminote}. The calculations are performed in a binned
$\Delta m^2,
\sin^2 2
\theta$ space and hence the bin edges are slightly irregular.

We see that the effective region at 90\% CL is from (2.80--3.20)  $\times 10^{-3}$ based on the off-axis data.  MINOS's errors are large on this
scale and contribute only a small amount; in any case some   of  the errors arising from the beam predictions ({\it e.g.}, total flux) are
correlated, so the improvement would be marginal at best.   If we were to {\it assume}
$\sin^2 2\theta$ is unity then a 90\% CL measurement would be 2.80-3.10 $\times 10^{-3}$ and a $1\sigma$ error would be about $0.1 \times 10^{-3}$. A 1\%
measurement of $\delta m^2$ would have an uncertainty of  $\pm .03 \times 10^{-3}$, about three times smaller than this work indicates. Of course
whether 
$\sin^2 2\theta$ is exactly unity is perhaps even more interesting than the precise value of $\Delta m^2$ and making such an assumption is unjustified from
the data.  A future paper will examine the $\nu_{\mu} \rightarrow \nu_e$ appearance channel including the effect of these uncertainties.


\end{document}